\documentclass[twocolumn,prl]{revtex4-2} 

\usepackage[T1]{fontenc}

\usepackage[scaled=0.92]{helvet}	

\usepackage[smallerops, cochineal, cochf]{newtxmath}
\usepackage{amsmath,amsfonts,bm}
\usepackage{graphicx}  
\usepackage{color}
\usepackage{extarrows}
\usepackage{enumitem}
\usepackage{empheq}
\usepackage{tensor}
\usepackage[cal=rsfso]{mathalfa}

\usepackage{esint}

\newcommand{\ud}{{\mathrm d}}

\newcommand{\wti}{\widetilde}

\newcommand{\tS}{\mbox{\tiny S}}

\newcommand{\SE}{\mbox{\tiny SE}}

\newcommand{\la}{\langle}
\newcommand{\ra}{\rangle}

\newcommand{\nl}{\nonumber \\}
\newcommand{\be}{\begin{equation}}
\newcommand{\ee}{\end{equation}}
\newcommand{\bsube}{\begin{subequations}}
\newcommand{\esube}{\end{subequations}}
\newcommand{\Eq}[1]{Eq.\,(\ref{#1})}

\newcommand{\Fig}[1]{Fig.\,\ref{#1}}

\newcommand{\RN}[1]{%
  \textup{\uppercase\expandafter{\romannumeral#1}}%
}

\allowdisplaybreaks[1]

\definecolor{darkblue}{RGB}{0, 56, 102}

\usepackage{hyperref}  
\hypersetup{hidelinks,
	colorlinks=true,
	allcolors=darkblue,
	pdfstartview=Fit,
	breaklinks=true
}

\begin{document}

\title{Stability of Quantum Systems beyond Canonical Typicality
}
\author{Yu Su}
\email{suyupilemao@mail.ustc.edu.cn}
\author{Zi-Fan Zhu} 
\author{Yao Wang}
\email{wy2010@ustc.edu.cn}
\author{Rui-Xue Xu}
\author{YiJing Yan}
\email{yanyj@ustc.edu.cn}
\affiliation{
Hefei National Research Center for Physical Sciences at the Microscale and Department of Chemical Physics, University of Science and Technology of China, Hefei, Anhui 230026, China
}

\date{\today}

\begin{abstract}
Involvement of the environment is indispensable for establishing the statistical distribution of system. We analyze the statistical distribution of a quantum system coupled strongly with a heat bath. This distribution is determined by tracing over the bath's degrees of freedom for the equilibrium system-plus-bath composite. The stability of system distribution is largely affected by the system--bath interaction strength. We propose that the quantum system exhibits a stable distribution only when its system response function in the frequency domain satisfies $\wti\chi(\omega = 0+)>0$. We show our results by investigating the non-interacting bosonic impurity system from both the thermodynamic and dynamic perspectives. Our study refines the theoretical framework of canonical statistics, offering insights into thermodynamic phenomena in small-scale systems.
\end{abstract}

\maketitle

\paragraph*{Introduction.}
%
In recent decades, researchers have delved into the fundamental principles of statistical physics by applying foundational quantum concepts \cite{Sch68,Tas981373,Gol06050403,Pop06754,Gel09051121,Sei16020601}. At the heart of quantum statistical mechanics lies the density operator of the canonical ensemble, expressed as:
\begin{align}\label{canonical}
  \rho_{\beta} = \frac{1}{Z_{\beta}}\exp\bigl( -\beta H_{\rm S} \bigr).
\end{align}
Here, $\beta\equiv1/(k_BT)$ denotes the inverse temperature, $H_{\rm S}$ is the system Hamiltonian, and $Z_{\beta}\equiv {\rm tr}\exp({-\beta H_{\rm S}})$ is the partition function. The typicality of the canonical distribution, as described in \Eq{canonical}, was established by Goldstein \textit{et al.}\ \cite{Gol06050403}. Their seminal study examines an isolated composite system consisting of a small system (S) and a large thermal bath (B). The principle of equal a priori probabilities results in the quantum state of the $\rm S+B$ composite system uniformly distributing over the Hilbert space within a narrow energy shell. Due to the high dimensionality of this uniform distribution, the central limit theorem is employed to show that for any $|\Psi\ra$ within the energy shell around $E$, 
\begin{align}\label{typicality}
  {\rm tr}_{\rm B}|\Psi\ra\la\Psi| \propto \sum_i \Omega_{\rm B}(E-\varepsilon_i)|\varepsilon_i\ra\la \varepsilon_i|,
\end{align}
where $\Omega_{\rm B}$ is the bath density of states and $\varepsilon_i$ is the $i$-th eigenvalue of system. Consequently, for a microscopic ensemble of $|\Psi\ra$, the system reduced density operator remains proportional to the right-hand side of \Eq{typicality}. Further assuming $\varepsilon_i\ll E$ and followed by the standard textbook derivation \cite{Cha87}, the canonical distribution is established.

It is noticed that the proof of \Eq{typicality} relies on the assumption of neglecting the system--bath interaction, i.e., $|V_{\rm SB}| \ll |H_{\rm S}|$. While this assumption holds true for systems in the thermodynamic limit, it may prove inadequate for small-scale systems subjected to significant environmental interactions. The exploration of statistical physics within the realm of small systems is crucial for advancing both theoretical frameworks \cite{Hil623182,Hil01111,Hil01273,Bin18,Gel09051121,Ile14032114} and practical applications. Hill's nanothermodynamics, as an early pioneering endeavor, studies the non-additive thermodynamic properties of such systems \cite{Hil623182}. Over time, this area of research has burgeoned, extending its relevance to diverse fields encompassing nano-science \cite{Cha1552,Qia12201}, surface physics \cite{Bed1840}, and biological physics \cite{Cao20eaaz4888}. Nevertheless, scant attention has been paid to how the environment affects the statistical distribution of system based on the foundational principles of quantum mechanics, especially when large system--bath integration exists.

In this Letter, we study the statistical distribution and the stability condition of quantum systems subject to significant environmental influence. Our theory starts with considering the adiabatic process of mixing the system and bath. The corresponding hybridization free energy \cite{Gon20154111,Gon20214115} characterizes the deviation of the system distribution from the canonical distribution. However, when the system--bath interaction strength surpass a critical value, the system may exhibit the instability. We propose that the quantum system exhibits a stable distribution only when its system response function satisfies 
\begin{align}
  \wti\chi(\omega = 0+) \equiv \int_0^\infty\!\chi(t)\,\ud t > 0.
\end{align}
We discuss our results by investigating the non-interacting bosonic impurity system. We show that violating the stability condition leads to the free energy being non-analytic at the critical point and thus being a non-physical observable. Furthermore, the stability condition also confirms whether the time evolution of system--plus--environment dynamics reaches equilibrium, which is deduced by examining the Routh--Hurwitz stability criterion for the exact equations of motion. Thorough this Letter, we set $\hbar\equiv 1$.

\begin{figure}[t]
  \centering
  \includegraphics[width=1.0\columnwidth]{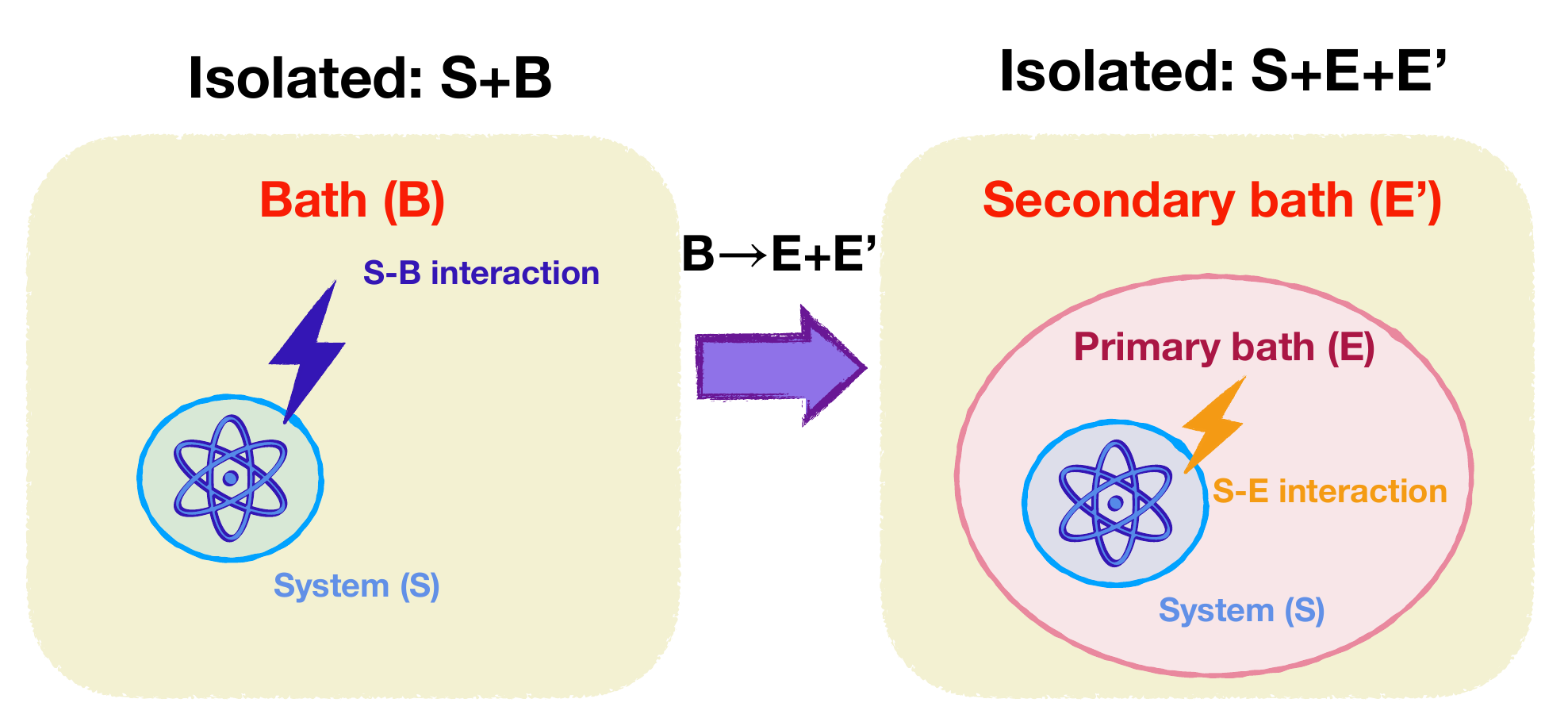}
  \caption{The left panel depicts the isolate composite: the system (S), the bath (B), and their interaction. We divide the bath B into the primary bath E and the secondary one $\rm E'$, shown in the right panel. The secondary bath remains the temperature given by $\beta=1/(k_BT)$ and the interactions of $\rm E'$ with others are ignored. The core system $\rm S+E$ satisfies the canonical typicality. }\label{fig1}
\end{figure}

\paragraph*{Setup.} 
Consider the isolated system--bath composite ($\rm S+B$) depicted in the left panel of \Fig{fig1}. We partition the thermal bath B into the primary environment $\mathrm{E}$ and the secondary one $\mathrm E'$. We stipulate that  $\rm E'$ remains its character as a thermal bath with inverse temperature $\beta$ and disregard the interactions between $\rm S+E$ and $\rm E'$. The validity of this partitioning relies on two key criteria: (i) the original bath $\mathrm B$ possesses an infinite number of degrees of freedom, and (ii) the ratio of the interaction between a system and a thermal bath to the system's internal energy diminishes with increasing system volume.Consequently, the state of $\mathrm S + \mathrm{E}$ is given by the canonical distribution, $\rho_{\rm S+E} = Z_{\rm S+E}^{-1}\exp (-\beta H_{\rm S+E})$. Here, the Hamiltonian reads $H_{\rm S+E} = H_{\rm S} + H_{\rm E} + V_{\rm SE}$ with $V_{\rm SE}$ being the system--environment interaction. 

The distribution of system S is characterized by the Hamiltonian of mean force $H_{\rm S}^\star (\beta)$, defined via \cite{Elc57161,Gel09051121,Hil11031110,New17032139,Mil18531,Naz18551}
\begin{align}\label{Heff}
  \rho_{\rm S} = {\rm tr}_{\rm E}\,\rho_{\rm S+E} \equiv \frac{1}{\mathcal Z_{\rm S}}\exp\big(-\beta H_{\rm S}^\star(\beta)\big),
\end{align}
where $\mathcal Z_{\rm S}\equiv Z_{\rm S+E}/Z_{\rm E} = {\rm tr}\exp(-\beta H_{\rm S}^\star(\beta))$ with $Z_{\rm E} = {\rm tr}_{\rm E}\exp(-\beta H_{\rm E})$.
The existence of interaction $V_{\rm SE}$ leads to $H_{\rm S}^\star(\beta) \neq H_{\rm S}$ in general. 
Note that the second--order perturbation master equation yields the high--temperature approximation \cite{Xu0211}, reading $\lim_{\beta\to0}H_{\rm S}^\star(\beta) \approx H_{\rm S} - \eta\hat Q^2$, with $\eta$ being the environment--induced reorganization energy and $\hat Q$ the system--interacting mode. 
There are many other approaches to analyzing $H_{\rm S}^\star(\beta)$; See \cite{Gel09051121,de202471,Che22064106,de235089}. We will not explore more in this work. 

\paragraph*{Hybridization free energy.}

Define the $\lambda$--augmented form of $\rm S+E$ composite Hamiltonian as $H_{\rm S+E}(\lambda) = H_{\rm S} + H_{\rm E} + \lambda V_{\rm SE}$
with $\lambda\in [0,1]$. 
The hybridization free energy reads \cite{Kir35300,Gon20154111,Gon20214115,Su24084104}
\begin{align}
  A_{\rm hyb}(\beta) &\equiv-\beta^{-1}\!\int_0^1\!\ud\lambda\,\frac{\partial}{\partial\lambda}\ln Z_{\rm S+E}(\beta;\lambda) \nl 
  &=  -\beta^{-1}(\ln \mathcal Z_{\rm S} - \ln Z_\beta),
\end{align}
with $Z_{\rm S+E}(\beta;\lambda) = {\rm Tr}_{}\exp(-\beta H_{\rm S+E}(\lambda))$ being the $\rm S+E$ space equilibrium partition function and $Z_\beta$ being the canonical partition function given by \Eq{canonical}. Here, ${\rm Tr}$ represents the trace over the S+E space. In fact, the difference between the thermal entropy defined via $-\partial A_{\rm hyb}/\partial T$ and the entanglement entropy of system serves as a measurement of the difference between $\rho_{\rm S}$ and $\rho_{\beta}$. We shall discuss this in the end of the Letter.

Using the relation, 
\begin{align*}
  {\rm Tr}\Big[\frac{\partial}{\partial\lambda}e^{\hat O(\lambda)}\Big]={\rm Tr}\Big[\frac{\partial \hat O(\lambda)}{\partial\lambda}e^{\hat O(\lambda)}\Big],
\end{align*}
we further have 
\begin{align}\label{Ahyb}
  A_{\rm hyb}(\beta)\! =\! \int_{0}^{1}\!\frac{{\ud}\lambda}{\lambda}\,{\rm Tr}\big[(\lambda V_{\rm SE})\rho_{\rm S+E}(\beta;\lambda)\big] \equiv \!\int_0^1\!\!\ud\lambda\,\la V_{\rm SE}\ra_\lambda,
\end{align}
with $\rho_{\rm S+E}(\beta;\lambda)=e^{-\beta H_{\rm S+E}(\lambda)}/Z_{\rm S+E}(\beta;\lambda)$. Equation (\ref{Ahyb}) interprets the hybridization free energy indeed as the integration of reversible work for the hybridization process, with the differential work being \cite{Gon20214115} $\delta w = {\rm Tr}[\delta H_{\rm S+E}(\lambda)\rho_{\rm S+E}(\beta;\lambda)] = \la V_{\rm SE}\ra_\lambda\delta\lambda$. On the other hand, \Eq{Ahyb} also provides us with a practical way to evaluate the hybridization free energy, just by calculating the mean value of the interaction, $\la V_{\rm SE}\ra_\lambda$, for all $\lambda\in[0,1]$.

\paragraph*{Stability condition.} To proceed, we consider the non-interacting bosonic impurity system (also named as the Brownian oscillator model). The Hamiltonian reads \cite{Cal83587,Xu037,Yan05187} 
\begin{align*}
  H_{\rm S} = \frac{\Omega_{\rm S}}{2}(\hat p^2+\hat q^2),\quad  H_{\rm E} = \sum_j\frac{\omega_j}{2}(\hat p_j^2+\hat x_j^2),
\end{align*}
and 
\begin{align*}
  V_{\rm SE} = \hat q\hat F = \hat q\sum_jc_j\hat x_j,
\end{align*}
where the dimensionless coordinates $\{\hat q,\hat x_j\}$ and momenta $\{\hat p,\hat p_j\}$ satisfy the Heisenberg commutation relations. 
The phase--space representation of $\rho_{\rm S}$ is of Gaussian distribution, characterized by the mean values \cite{Cal83587,Xu037,Yan05187} , $\la\hat q\ra = \la\hat p\ra = 0$, and the variances, $\la\delta\hat p^2\ra$ and $\la\delta\hat q^2\ra$, with $\la\cdots\ra \equiv {\rm Tr}\,(\cdots\rho_{\rm S+E})$. 
Then we obtain the Hamiltonian of mean force \cite{Xu037,Yan05187,Mil18531}, $H^\star_{\rm S}(\beta) = \frac{1}{2}\Omega_{\text{eff}}(\beta)(\hat p^2 + \hat q^2)$
with $\Omega_{\text{eff}}(\beta) = 2\beta^{-1}{\rm arcoth}(2\sqrt{\la\delta p^2\ra\la \delta q^2\ra})$ being the effective system frequency. 
The fluctuation--dissipation theorem yields \cite{Wei12,Xu037}
\begin{align}
    \begin{split}
        \la\delta\hat q^2\ra &= \frac{1}{\beta}\wti \chi(0) + \frac{2}{\beta}\sum_{n=1}^\infty\wti \chi(i\varpi_n),\\
        \la\delta\hat p^2\ra &= \frac{1}{\beta\Omega_{\rm S}} + \frac{2}{\beta\Omega_{\rm S}}\sum_{n=1}^\infty\big[ 1 - \Omega_{\rm S}^{-1}\varpi_n^2\wti\chi(i\varpi_n) \big],
    \end{split}
\end{align}
with $\{\varpi_n = 2\pi n/\beta\}$ being the Matsubara frequencies. Here, we define the system response function $\chi(t) \equiv i\la[\hat q(t),\hat q(0)]\ra$ and its frequency domain $\wti \chi(\omega)$ with denoting $\wti f(\omega) \equiv \int_0^\infty\!\ud t\,e^{i\omega t}f(t)$ for any $f(t)$. 
By exploring the Heisenberg equations for $\hat q(t)$ and $\hat p(t)$, we obtain 
\begin{align}\label{chi}
    \wti \chi(\omega) = \frac{\Omega_{\rm S}}{\Omega_{\rm S}^2 - \omega^2 - \Omega_{\rm S}\wti\phi_{\rm E}(\omega)},
\end{align}
where $\phi_{\rm E}(t) \equiv \sum_j c_j^2\sin(\omega_jt) = i\la[\hat F_{\rm E}(t),\hat F_{\rm E}(0)]\ra_{\rm E}$ is the bare--environment response function. Specifically, $\hat F_{\rm E}(t) \equiv e^{iH_{\rm E}t}\hat Fe^{-iH_{\rm E}t}$ and $\la\cdots\ra_{\rm E} \equiv {\rm tr}_{\rm E}(\cdots e^{-\beta H_{\rm E}}/Z_{\rm E})$.
%
With the same procedure, we have the hybridization free energy, reading \cite{Gon20214115}
\begin{align}
    A_{\rm hyb}(\beta) = -\frac{1}{\beta}\vartheta(0) - \frac{2}{\beta}\sum_{n=1}^\infty\vartheta(\varpi_n),
\end{align}
where $\vartheta(\omega)$ is the free-energy spectral function, evaluated via 
\begin{align}\label{vartheta}
    \vartheta(\omega) &= {\rm Re}\!\int_0^1\!\ud\lambda\,\lambda\wti\phi_{\rm E}(i\omega)\wti\chi(i\omega;\lambda)\nl 
    &= \frac{1}{2}\,{\rm ln}\bigg| \frac{\Omega_{\rm S}^2 + \omega^2}{\Omega_{\rm S}^2 + \omega^2 - \Omega_{\rm S}\wti\phi_{\rm E}(i\omega)} \bigg|,
\end{align}
with $\wti \chi(\omega;\lambda)$ defined through the $\lambda$--augmented Hamiltonian, i.e., $V_{\SE}\to\lambda V_{\SE}$.

\begin{figure}[htp]
  \centering
  \includegraphics[width=0.95\columnwidth]{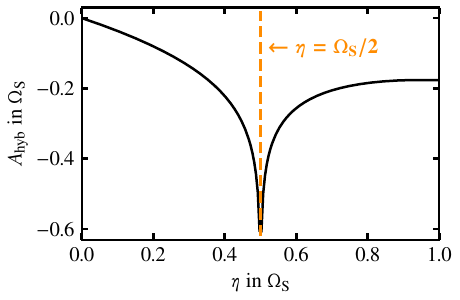}
  \caption{The $\eta$--dependence of the hybridization free energy $A_{\rm hyb}(\beta)$ for the non-interacting bosonic impurity system. The environment response is modeled using the Drude spectral function, $\wti\phi_{\rm E}(\omega) = \eta\gamma/(\gamma-i\omega)$. The parameters are given by $\gamma = 2\Omega_{\rm S}$ and $k_BT = 0.2\Omega_{\rm S}$. }\label{fig2}
\end{figure}

Define the environment--induced reorganization energy as \cite{Wei12,Cal83587,Yan05187} $\eta \equiv \wti\phi_{\rm E}(0)/2 = \sum_jc_j^2/(2\omega_j)$, which serves as a measure for the system--environment interaction strength. We see from \Eq{vartheta} that $\vartheta(\omega)$ diverges at $\omega=0$ when the environment--induced reorganization energy reaches $\Omega_{\rm S}/2$. Consequently, the hybridization free energy is not analytic in the $\eta$--space. \Fig{fig2} depicts the $\eta$--dependence of $A_{\rm hyb}$. We see that the orange dashed line $\eta = \Omega_{\rm S} / 2$ separates $A_{\rm hyb}$ into two different regions. However, we shall show that only within $\eta\in(0,\Omega_{\rm S}/2)$ the system is stable and the free energy is physically observed. 

We first note that, at thermal equilibrium, the response function $\chi(t)$ must be finite-time supported \cite{Wei12}. In detail, there exists a constant $K>0$, such that for all $t>0$, we have $|\chi(t)| \leq K.$ This leads to the Fourier transformation $\wti\chi(\omega)$ being analytic in the upper--half plane. By exploring the Cramers-Kr\"{o}nig relation and the fluctuation--dissipation theorem, we obtain \cite{Wei12}
\begin{align}\label{eq11}
  \lim_{\omega\to 0+}\wti\chi(\omega) = \frac{1}{\pi}{\rm Im}\!\int_{-\infty}^\infty\!\!\ud\omega\,\frac{\wti\chi(\omega)}{\omega} > 0.
\end{align} 
Substituting \Eq{chi} into \Eq{eq11} yields $\Omega_{\rm S} - 2\eta > 0$. Similarly, for the bare--environment response, we have $\wti\phi_{\rm E}(0+) = 2\eta > 0$, which is consistent with the definition of the reorganization energy. The finite supported property is general in a dissipative quantum system. For a system at thermal equilibrium, the correlation functions must follow the asymptotically uncorrelated statistics, $\lim_{t\to\infty}\la\hat A(t)\hat B(0)\ra = \la\hat A\ra\la\hat B\ra$, for arbitrary observables \cite{Wei12,Cal83587,Xu037,Yan05187}. 

\begin{center}
    \begin{figure*}[htp]
      \centering
      \includegraphics[width=1\textwidth]{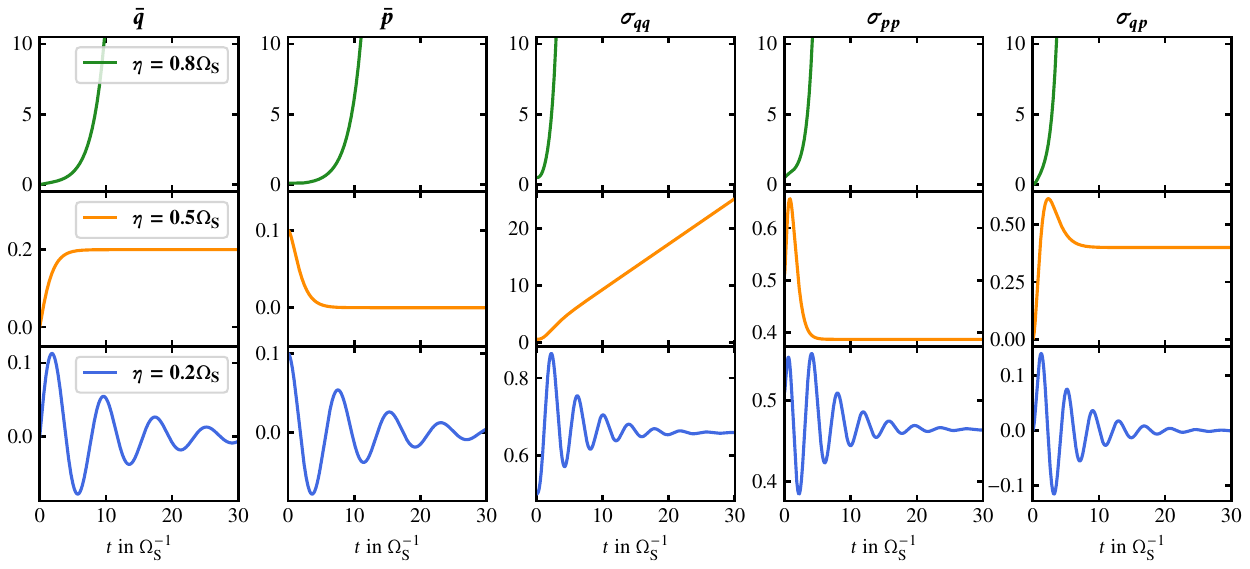}
      \caption{The time evolution of $\bar q(t) \equiv {\rm tr}[\hat q\rho_{\rm S}(t)]$, $\bar p(t) \equiv {\rm tr}[\hat p\rho_{\rm S}(t)]$, $\sigma_{qq}(t) \equiv {\rm tr}[\hat q^2\rho_{\rm S}(t)]$, $\sigma_{pp}(t) \equiv {\rm tr}[\hat p^2\rho_{\rm S}(t)]$, and $\sigma_{qp}(t) \equiv {\rm tr}[\hat q\hat p\rho_{\rm S}(t)]$ at $\eta = 0.2$, $0.5$, and $0.8\Omega_{\rm S}$. The Drude spectral function is adopted for modeling environment. The parameters are given by $\gamma=2\Omega_{\rm S}$ and $k_BT = 0.2\Omega_{\rm S}$.}\label{fig3}
    \end{figure*}
  \end{center}

Furthermore, we analyze the stability dynamics for the non-interacting bosonic impurity system, using the Routh--Hurwitz criterion \cite{Rao19}. The equations of motion are derived based on the hierarchical equations of motion (HEOM), an exact formalism for open quantum systems \cite{Tan906676,Yan04216,Xu05041103,Yan16110306,Tan20020901}. For simplicity, we consider the Drude environment case, $\wti\phi_{\rm E}(\omega) = 2\eta\gamma/(\gamma-i\omega)$. By using the HEOM, we obtain the equations of motion for mean values, reading \cite{Yan16110306}
\begin{equation}\label{EOM}
  \begin{split}
    \dot{\bar q} &= \Omega_{\rm S}\bar p,\\
    \dot{\bar p} &= -\Omega_{\rm S}\bar q - \varphi - \sum_{n=1}^\infty\theta_n,\\
    \dot\varphi &= -\gamma\varphi - 2\eta\gamma\bar q, \\
    \dot\theta_n &= -\varpi_n\theta_n.
  \end{split}
\end{equation}
Here, $\bar q(t)\equiv {\rm tr}_{}[\hat q\rho_{\rm S}(t)]$, $\bar p(t)\equiv {\rm tr}_{}[\hat p\rho_{\rm S}(t)]$, and $\{\varphi(t), \theta_n(t)\}$ are auxiliary variables characterizing the environmental influences. Utilizing the Hurwitz theorem \cite{Rao19}, we find that the dynamics described by \Eq{EOM} is stable if and only if 
\begin{align*}
  f(z) = z^3 + \gamma z^2 + \Omega_{\rm S}^2 z + (\Omega_{\rm S} - 2\eta)\gamma\Omega_{\rm S}
\end{align*}
is a Hurwitz polynomial, i.e., all of its roots should be distributed in the left--half complex plane \cite{Rao19}. By applying the Routh--Hurwitz criterion \cite{Rao19} on the corresponding Hurwitz matrix, one can obtain the stability conditions being $\Omega_{\rm S}-2\eta>0$ and $\eta>0$. In other words, the system cannot reach thermal equilibrium when the stability conditions are violated. In \Fig{fig3}, we plot the time evolution results of the mean values and variances of the system Gaussian distribution at different reorganization energies. We see that all the quantities diverge at $\eta = 0.8\Omega_{\rm S}$, while they reach equilibrium at $\eta = 0.2\Omega_{\rm S}$. For the critical case, $\eta = 0.5\Omega_{\rm S}$, the dynamics behaves like a random walk: The variance of coordinate $\sigma_{qq}(t)$ is proportional to $t$ in long time, but other quantities converge. This is because the effective system frequency is reorganized to zero at the critical point. 

The above discussions reveal that the stability of the system distribution is largely affected by the system--bath interaction strength. One key criterion to ensure the system reaches thermal equilibrium is that $\wti\chi(0+)$ must be greater than zero. For this reason, we usually add the total system--plus--environment Hamiltonian a reorganization energy term, $H_{\rm re} = \eta\hat q^2$, in order to maintain the dynamic stability of the system \cite{Cal83587,Xu0211,Xu037,Yan05187}. However, the reorganization energy would impose an additional positive contribution to the hybridization free energy \cite{Su24084104}, $A_{\rm hyb}^{\rm re} = 2\int_0^1\!\ud\lambda\,\lambda\eta\la\hat q^2\ra_\lambda$.

\paragraph*{Subdivision potential.} Let us turn back to the distribution under the stability conditions. To identify the deviation of $\rho_{\tS}$ from the canonical distribution $\rho_\beta$, an acceptable choice is the subdivision potential $\mathcal E$ \cite{Hil623182,de202471},
\begin{align}\label{subdivision}
  \mathcal E \equiv E_{\rm S} - \la H_{\rm S}^\star(\beta)\ra,
\end{align} 
with $E_{\rm S} \equiv -\frac{\partial}{\partial\beta}\ln \mathcal Z_{\rm S}$. The subdivision potential is firstly introduced by Hill to accommodate the non-extensive thermodynamic behavior of small systems \cite{Hil623182,Hil01111,Hil01273}. 
%
The subdivision potential approaches to zero when (i) $H_{\text{eff}}(\beta)$ varies barely with the temperature or (ii) the system S is large enough such that the interaction term is negligible. 
%
%
%
For practical use, we recast \Eq{subdivision} as 
\begin{align}\label{entropy}
  \mathcal E = T(S_{\rm therm} - S_{\rm ent}),
\end{align}
where we define the thermodynamic entropy, $S_{\rm therm} \equiv -\frac{\partial }{\partial T}A_{\rm therm}$
with $A_{\rm therm} = -k_BT\ln \mathcal Z_{\rm S}$, and the entanglement entropy (or von Neumann entropy), $S_{\rm ent} \equiv -k_B{\rm tr}_{\rm S}(\rho_{\rm S}\ln\rho_{\rm S})$. 
%
%
%
%
While $S_{\rm ent}$ characterizes the local information regarding solely the system, the hybridization entropy $S_{\rm therm}$ encompasses both the local and non-local contributions arising from the system--bath hybridization.
These two entropies may exhibit disparate behaviors \cite{Le07126801} in the context of quantum impurity systems, leading to a notable subdivision potential and a markedly non-canonical distribution. 
The condition $\mathcal E=0$ serves as a criterion for determining whether the system conforms to a canonical distribution, as at this juncture, the significance of non-local entanglement information involving the environment diminishes.

\begin{figure}[htp]
  \centering
  \includegraphics[width=0.95\columnwidth]{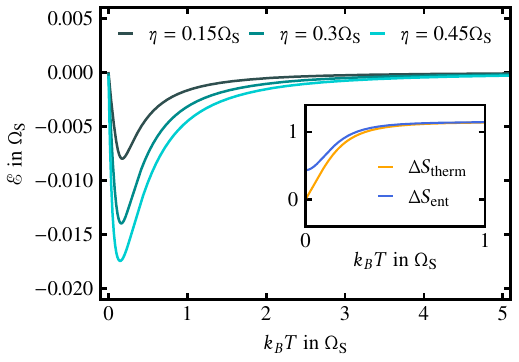}
  \caption{The main panel presents the temperature dependence of the subdivision potential for $\eta = 0.15,\ 0.3,$ and $0.45 \Omega_{\rm S}$. In the inset panel, $\Delta S_{\rm therm} \equiv S_{\rm therm} - S_\beta$ and $\Delta S_{\rm ent} \equiv S_{\rm ent} - S_\beta$ as functions of temperature are depicted specifically at $\eta = 0.3\Omega_{\rm S}$, where $S_\beta$ is the entropy defined via \Eq{canonical}. The bath response is modeled using the Drude spectral function. We set $\gamma = 2\Omega_{\rm S}$. }\label{fig4}
\end{figure}

Figure \ref{fig4} depicts the temperature dependence of the subdivision potential at various values of $\eta$. Notably, we observe that for a fixed $\eta$, the energy $\mathcal E$ diminishes as the temperature tends towards either zero or infinity. This trend stems from the behavior of the difference between the thermodynamic and entanglement entropies, which is finite at zero temperature and diminishes at high temperature. Moreover, the subdivision potential minimizes around $k_BT \sim 0.2\Omega_{\rm S}$, showing little sensitivity to changes in $\eta$. Additionally, the magnitude of $\mathcal E$ escalates as the interaction strength between the system and the bath increases.

One may generalize our discussion to other ensemble setups. For the grand canonical case, the particle number operator is additive \cite{de202471}, i.e., $\hat N_{\rm S+\rm E} = \hat N_{\rm S} + \hat N_{\rm E}$. 
The equilibrium state of the $\rm S+E$ composite then follows $\rho_{\rm S+E} \propto \exp{[-\beta(H_{\rm S+E}-\mu \hat N_{\rm S+E})]}$, with $\mu$ being the chemical potential.
The Hamiltonian of mean force is then constructed by replacing $H_{\rm S}^\star$ with $H_{\rm S}^\star(\beta,\mu) -\mu\hat N_{\rm S}$ in \Eq{Heff}. 
The subdivision potential is still defined by $E_{\rm S} - \la H_{\rm S}^\star(\beta,\mu)\ra$.

\paragraph*{Summary.}

To summary, this Letter investigates the statistical distribution of a quantum system coupled with a heat bath. By scrutinizing the equilibrium state of the system-bath composite and tracing over the bath's degrees of freedom, we unveil the underlying distribution. Our investigation entails assessing the free-energy change throughout the system-bath hybridization process, offering crucial insights into the system's stability. Furthermore, we explore the analytic properties of the free-energy spectral function to characterize the quantum system's stability. By exploring the exact equations of motion, we obtain that the key criterion to ensure the system both dynamic and thermodynamic stable is that $\wti\chi(0+)$ must larger that zero. Introducing the subdivision potential as a measure, we quantify deviations from the canonical distribution, elucidating disparities between thermodynamic and entanglement entropies. This work refines the theoretical underpinnings of canonical statistics, providing a nuanced understanding of thermodynamic phenomena in small-scale systems, with implications for a range of physical contexts.

\vspace{1em}

Support from the Ministry of Science and Technology of China, Grant No.\ 2017YFA0204904, and the National Natural Science Foundation of China, Nos.\ 22103073, 22173088 and  22373091 are gratefully acknowledged. This research is partially motivated by the 1st ``Question and Conjecture'' activity supported by the Top Talent Training Program 2.0 for undergraduates.



\begin{thebibliography}{10}

  \bibitem{Sch68}
  E.~Schrodinger,
  \newblock {\em Statistical Thermodynamics},
  \newblock 1968.
  
  \bibitem{Tas981373}
  H.~Tasaki,
  \newblock Phys. Rev. Lett. {\bf 80}, 1373 (1998).
  
  \bibitem{Gol06050403}
  S.~Goldstein, J.~L. Lebowitz, R.~Tumulka, and N.~Zangh{\`i},
  \newblock Phys. Rev. Lett. {\bf 96}, 050403 (2006).
  
  \bibitem{Pop06754}
  S.~Popescu, A.~J. Short, and A.~Winter,
  \newblock Nature Phys {\bf 2}, 754 (2006).
  
  \bibitem{Gel09051121}
  M.~F. Gelin and M.~Thoss,
  \newblock Phys. Rev. E {\bf 79}, 051121 (2009).
  
  \bibitem{Sei16020601}
  U.~Seifert,
  \newblock Phys. Rev. Lett. {\bf 116}, 020601 (2016).
  
  \bibitem{Cha87}
  D.~Chandler,
  \newblock {\em Introduction to Modern Statistical Mechanics},
  \newblock Oxford University Press, New York, 1st edition edition, 1987.
  
  \bibitem{Hil623182}
  T.~L. Hill,
  \newblock J. Chem. Phys. {\bf 36}, 3182 (1962).
  
  \bibitem{Hil01111}
  T.~L. Hill,
  \newblock Nano Lett. {\bf 1}, 111 (2001).
  
  \bibitem{Hil01273}
  T.~L. Hill,
  \newblock Nano Lett. {\bf 1}, 273 (2001).
  
  \bibitem{Bin18}
  F.~Binder, L.~A. Correa, C.~Gogolin, J.~Anders, and G.~Adesso, editors,
  \newblock {\em Thermodynamics in the Quantum Regime: Fundamental Aspects and New Directions}, volume 195 of {\em Fundamental Theories of Physics},
  \newblock Springer International Publishing, Cham, 2018.
  
  \bibitem{Ile14032114}
  J.~{Iles-Smith}, N.~Lambert, and A.~Nazir,
  \newblock Phys. Rev. A {\bf 90}, 032114 (2014).
  
  \bibitem{Cha1552}
  R.~V. Chamberlin,
  \newblock Entropy {\bf 17}, 52 (2015).
  
  \bibitem{Qia12201}
  H.~Qian,
  \newblock J. Biol. Phys. {\bf 38}, 201 (2012).
  
  \bibitem{Bed1840}
  D.~Bedeaux and S.~Kjelstrup,
  \newblock Chem. Phys. Lett. {\bf 707}, 40 (2018).
  
  \bibitem{Cao20eaaz4888}
  J.~Cao, R.~J. Cogdell, D.~F. Coker, H.-G. Duan, J.~Hauer, U.~Kleinekath{\"o}fer, T.~L.~C. Jansen, T.~Man{\v c}al, R.~J.~D. Miller, J.~P. Ogilvie, V.~I. Prokhorenko, T.~Renger, H.-S. Tan, R.~Tempelaar, M.~Thorwart, E.~Thyrhaug, S.~Westenhoff, and D.~Zigmantas,
  \newblock Sci. Adv. {\bf 6}, eaaz4888 (2020).
  
  \bibitem{Gon20154111}
  H.~Gong, Y.~Wang, H.-D. Zhang, Q.~Qiao, R.-X. Xu, X.~Zheng, and Y.~Yan,
  \newblock J. Chem. Phys. {\bf 153}, 154111 (2020).
  
  \bibitem{Gon20214115}
  H.~Gong, Y.~Wang, H.-D. Zhang, R.-X. Xu, X.~Zheng, and Y.~Yan,
  \newblock J. Chem. Phys. {\bf 153}, 214115 (2020).
  
  \bibitem{Elc57161}
  E.~W. Elcock and P.~T. Landsberg,
  \newblock Proc. Phys. Soc. B {\bf 70}, 161 (1957).
  
  \bibitem{Hil11031110}
  S.~Hilt, B.~Thomas, and E.~Lutz,
  \newblock Phys. Rev. E {\bf 84}, 031110 (2011).
  
  \bibitem{New17032139}
  D.~Newman, F.~Mintert, and A.~Nazir,
  \newblock Phys. Rev. E {\bf 95}, 032139 (2017).
  
  \bibitem{Mil18531}
  H.~J.~D. Miller,
  \newblock in {\em Thermodynamics in the Quantum Regime}, edited by F.~Binder, L.~A. Correa, C.~Gogolin, J.~Anders, and G.~Adesso, volume 195, pages 531--549, Springer International Publishing, Cham, 2018.
  
  \bibitem{Naz18551}
  A.~Nazir and G.~Schaller,
  \newblock in {\em Thermodynamics in the Quantum Regime: Fundamental Aspects and New Directions}, edited by F.~Binder, L.~A. Correa, C.~Gogolin, J.~Anders, and G.~Adesso, Fundamental Theories of Physics, pages 551--577, Springer International Publishing, Cham, 2018.
  
  \bibitem{Xu0211}
  R.~Xu and Y.~Yan,
  \newblock J. Chem. Phys. {\bf 116}, 11 (2002).
  
  \bibitem{de202471}
  R.~{de Miguel} and J.~M. Rub{\'i},
  \newblock Nanomaterials {\bf 10}, 2471 (2020).
  
  \bibitem{Che22064106}
  T.~Chen and Y.-C. Cheng,
  \newblock J. Chem. Phys. {\bf 157}, 064106 (2022).
  
  \bibitem{de235089}
  R.~{de Miguel},
  \newblock J. Phys. Chem. B {\bf 127}, 5089 (2023).
  
  \bibitem{Kir35300}
  J.~G. Kirkwood,
  \newblock J. Chem. Phys. {\bf 3}, 300 (1935).
  
  \bibitem{Su24084104}
  Y.~Su, Y.~Wang, R.-X. Xu, and Y.~Yan,
  \newblock J. Chem. Phys. {\bf 160}, 084104 (2024).
  
  \bibitem{Cal83587}
  A.~O. Caldeira and A.~J. Leggett,
  \newblock Physica A: Statistical Mechanics and its Applications {\bf 121}, 587 (1983).
  
  \bibitem{Xu037}
  R.~Xu, Y.~Mo, P.~Cui, S.-H. Lin, and Y.~Yan,
  \newblock in {\em Advanced Topics in Theoretical Chemical Physics}, edited by W.~N. Lipscomb, J.~Maruani, S.~Wilson, J.~Maruani, R.~Lefebvre, and E.~J. Br{\"a}ndas, volume~12, pages 7--40, Springer Netherlands, Dordrecht, 2003.
  
  \bibitem{Yan05187}
  Y.~Yan and R.~Xu,
  \newblock Annu. Rev. Phys. Chem. {\bf 56}, 187 (2005).
  
  \bibitem{Wei12}
  U.~Weiss,
  \newblock {\em Quantum Dissipative Systems},
  \newblock Wspc, New Jersey, 4th ed. edition edition, 2012.
  
  \bibitem{Rao19}
  J.~R. Raol and R.~Ayyagari,
  \newblock {\em Control Systems: Classical, Modern, and AI-Based Approaches},
  \newblock CRC Press, Boca Raton, 2019.
  
  \bibitem{Tan906676}
  Y.~Tanimura,
  \newblock Phys. Rev. A {\bf 41}, 6676 (1990).
  
  \bibitem{Yan04216}
  Y.-a. Yan, F.~Yang, Y.~Liu, and J.~Shao,
  \newblock Chem. Phys. Lett. {\bf 395}, 216 (2004).
  
  \bibitem{Xu05041103}
  R.-X. Xu, P.~Cui, X.-Q. Li, Y.~Mo, and Y.~Yan,
  \newblock J. Chem. Phys. {\bf 122}, 041103 (2005).
  
  \bibitem{Yan16110306}
  Y.~Yan, J.~Jin, R.-X. Xu, and X.~Zheng,
  \newblock Front. Phys. {\bf 11}, 110306 (2016).
  
  \bibitem{Tan20020901}
  Y.~Tanimura,
  \newblock J. Chem. Phys. {\bf 153}, 020901 (2020).
  
  \bibitem{Le07126801}
  K.~Le~Hur, P.~{Doucet-Beaupr{\'e}}, and W.~Hofstetter,
  \newblock Phys. Rev. Lett. {\bf 99}, 126801 (2007).
  
  \end{thebibliography}

\end{document}